# Shift of Fermi level by substitutional impurity-atom doping in diamond and cubic- and hexagonal-boron nitrides

Koji Kobashi

Shinko Research Co. Ltd., Division of Industrial Strategy Information
2-2-4 Wakino-hama Kaigan-dori, Chuo-ku, Kobe 651-0073, Japan
(E-mail: paris16eme2001@ybb.ne.jp)

Abstract

   The density of states and the band diagrams were computed for diamond, cubic boron nitrde (cBN), and hexagonal boron nitride (hBN) using a Korringa-Kohn-Rostoker (KKR) scheme to investigate the shift of the Fermi level by impurity-atom doping below 10 at.%. The dopant atoms were B and N for diamond, Be, Si, and C for cBN, and Be and C for hBN. It was found that the Fermi level was located at the valence band maximum or the conduction band minimum in the following seven cases: (i) the B concentration was 0.3 at.% in B-doped diamond, (ii) the N concentration was 0.4 at.% in N-doped diamond, (iii) the concentration of Be substituting B was 0.9 at.% in cBN, (iv) the concentration of Si substituting B was 0.3 at.% in cBN, (v) the concentration of C substituting B was 0.3 at.% in cBN, (vi) the concentration of C substituting N was 0.9 at.% in cBN, and (vii) the concentration of Be substituting B was ~2 at.% in hBN. Each of these values indicates the critical dopant concentration for semiconductor-to-metal transition. In B-doped diamond, it serves a measure for the occurrence of superconductivity at low temperature.

I. Introduction

   Diamond is intrinsically an insulator with an indirect bandgap energy of 5.47 eV [1, 2], but becomes a *p*-type semiconductor by boron (B)-doping with an acceptor level at 0.37 eV above the valence band maximum (VBM) in the center of the Brillouin zone (BZ), the Γ-point. B-doped diamonds exhibit metallic conduction when the B concentration, $n_B$, is above $3 \times 10^{20}$ /cm$^3$ (0.17 at.%) [3] or $5 \times 10^{20}$ /cm$^3$ (0.28 at.%) [4, 5]. Since diamond is a wide bandgap semiconductor, the research and development has been focused on its high temperature characteristics. In 2004, Ekimov *et al.* [6-9] found, however, that heavily B-doped diamonds ($n_B \geq 2.8 \pm 0.5$ at.%, or $5.0 \times 10^{20}$ /cm$^3$, and the carrier concentration $\geq 2 \times 10^{21}$ /cm$^3$), synthesized by the high pressure-high temperature (HPHT) method, exhibited zero-electrical resistance at 2 K. This finding stimulated extensive studies of diamond superconductivity by many groups [10-36]. Many of these studies used diamond films because the microwave plasma chemical vapor deposition (CVD) technology had been well established for controlled B-doping in high-quality diamond films [37]. One of the issues on diamond superconductivity was that the onset temperatures were different in (111), (100), and (110) oriented diamonds, even though the values of $n_B$, determined by secondary ion mass spectroscopy (SIMS), were the same [33, 38]. This issue was later solved by a nuclear magnetic resonance (NMR) study of diamond doped with $^{11}$B isotope of $^{10}$B [14, 23]. It was concluded that among B atoms (i) substituted with carbon (C) atoms, (ii) present in interstitial positions, or (iii) in B-H complex states, only substitutional B atoms were concerned with the superconductivity. Indeed, by estimating substitutional B concentrations, $n_B^s$, from NMR signals, the onset temperatures, $T_c$(onset), of superconductivity in the previous works using (111), (100), and (110) oriented diamonds were able to be plotted on a single smooth curve as a function of log $n_B^s$. In many of later works, the superconducting properties were studied on the basis of carrier concentrations, not $n_B$, to avoid such an orientational effect. In this regard, Ref. [16] is of great use to know the relations between the carrier concentrations, the B concentrations, and the onset temperatures for (111), (100), and (110) oriented films. This work concluded that $n_B^s \geq 3 \times 10^{20}$ /cm$^3$ was necessary for superconductivity to occur. The critical carrier concentration for superconductivity-to-insulator, evaluated by Hall measurements, was $4 \times 10^{20}$ /cm$^3$; it was higher than $n_B^s$. The band structures around the Γ-point in (111) oriented homoepitaxial diamond films with different B concentrations were observed by soft x-ray angle-resolved photoemission spectroscopy (SXARPES) [19, 29, 38]. An analysis, in reference to a theoretical band structure, showed that a diamond film of $n_B = 2.88 \times 10^{20}$ /cm$^3$ (0.16 at.%) did not become superconducting; the Fermi level was about 0.10 eV *above* the VBM. On the other hand, diamond films of $n_B = 1.18 \times 10^{21}$ /cm$^3$ (0.67 at.%) and $8.37 \times 10^{21}$ /cm$^3$ (4.73 at.%) became superconducting at low temperature; the Fermi level was 0.2 and 0.4 eV *below* the VBM, respectively. Moreover, a photoemission spectroscopy [19] suggested that the Fermi levels of B-doped diamonds were located not on the acceptor band but below the VBM in case that the diamonds became superconducting. These experimental results suggest that the $n_B$-value at which the Fermi level is at the VBM serves a good measure for superconductivity to take place at low temperature. Hence the determination of $n_B$ at which the Fermi level is at the VBM by simulation is the major objective in the present paper on B-doped diamond. Theoretical study of diamond superconductivity has been a hot topic since Ekimov *et al.* [6], and the superconductivity is widely understood by the theory of Bardeen, Cooper, and Schrieffer (the BCS theory) [24, 39], where the carriers are the electrons in the unfilled valence band, *i.e.*, the holes around the Γ-point. Regarding the research of superconductivity in diamond up to 2009, the reader can refer to excellent reviews by Bustarret [12, 40], Takano [15, 38], and Nesladek [41].

For the *n*-type dopant of diamond, nitrogen (N) is considered to be the best candidate in view of the atomic radius similar to C (N: 0.75 Å, C: 0.77 Å), but it actually creates a deep donor level at approximately 1.7 eV below the conduction band minimum (CBM) due to the Jahn-Teller (JT) effect so that virtually no electron is excited to the conduction band at the ambient temperature. Knowing all those things, the density of states (DOS) and the band diagrams of N-doped diamond were computed in the present paper to compare with similar studies in the past [42-45]. In experimental research, phosphorous (P) is successfully used for an *n*-type dopant, but this dopant atom is out of scope in the present paper because the local lattice distortion, which seemed to be significant around the P atoms, was not taken into account in the present work [46-48].

Cubic boron nitride (cBN) also is a good insulator with an indirect bandgap energy of 6.3 eV between the VBM at the Γ-point and the CBM at the X-points [49], or 6.36 eV according to the luminescence excitation spectroscopy [50]. Ultraviolet (UV) photoluminescence measurements of single crystal cBN determined the bandgap to be 6.25 eV [51]. A wide range of material properties of cBN is compiled in Refs. [49, 52]. Cubic BN can be synthesized by either the HPHT method or CVD, but cBN films grown by CVD include other BN phases such as hexagonal boron nitride (hBN) [53, 54], while bulk cBN grown at HPHT is usually a single phase [51]. It is known that cBN becomes a *p*-type semiconductor by beryllium (Be)-doping [49, 55] and an *n*-type semiconductor by silicon (Si) doping [49]. This is supported by the formation of a rectifying and light emitting *p-n* junction using B- and Si-doped cBN [56, 57]. It is also reported that the electrical resistivity of Si-doped cBN films markedly decreases for the Si concentration $n_{Si} \geq 1.5$ at.%, though the resistivity is still high [53]. Doping of C and other elements in cBN has also been studied (see references cited in Refs. [49, 55]). Since BN has various possibilities of dopant atom substitution, *i.e.*, either B or N site, or both B and N sites, it is necessary to specify both dopant atom and doping site(s) to describe the doping effect on the electronic state. In the present paper, Be, Si, and C were used as the substitutional dopant atoms at the B and N sites of cBN, and the shift of the Fermi level was computed for different dopant concentrations.

Hexagonal BN [52, 58] has a number of polytypes that have been extensively studied theoretically by Kobayashi [59] and other researchers [60-62]. It consists of hexagonal B-N layers, and there are a variety of ways in stacking them along the c-direction of the hexagonal axis. Like graphite, the van der Waals force works between the B-N layers so that the energy differences among different stackings are quite small [59]. In the present paper, only a most common, so-called A-B stacking structure of hexagonal BN is considered, and will be referred to as hBN hereafter. Recently, Watanabe and co-workers found an intense UV luminescence [63, 64] as well as a UV laser light emission [65] at 215 nm by cathodoluminescence (CL) from bulk single-crystal hBN synthesized at HPHT and at room temperature as well [51, 66]. These works brought a renewed interest in controversial issues on intrinsic properties of hBN in both theory and experiment, *e.g.*, bandgap energy, direct *vs.* indirect transition, and the transition points in the band diagram [50, 67], mainly because high quality single crystals had not been available for accurate measurements. A current understanding on hBN, based on sophisticated theories [68] and experiments using bulk single crystal hBN [51, 66], is that it has an indirect bandgap with an energy of 5.7 eV and the binding energy of exciton is as deep as 0.72 eV [69]. A recent photoluminescence study [70], however, claimed that the bandgap energy is 6.4 eV, and the binding energy of exciton is greater than 0.38 eV, while a luminescence excitation spectroscopy [50] determined the quasi-direct bandgap to be 5.96 eV. Thus, the bandgap energy and the exciton states are still under debate, and intensive studies are ongoing [69, 71-75]. In view of theory, there are a number of challenging issues to be solved on hBN: determining the most stable stacking structure and including the van der Waals force in the theory are the fundamental issues, and identifying the indirect transition points in the band diagram is a difficult issue because of the unique feature of the band diagram that the valence band is partly flat [52, 68, 76], and the band diagram is dependent on the theoretical technique used. A number of theoretical studies of undoped hBN are present, but the above issues have not yet been understood unequivocally. In the present paper, the shift of the Fermi level was computed using Be and C as substitutional dopants at the B and N sites of hBN.

In response to the recent progress in experimental research, the electronic states of doped and undoped diamond, cBN, and hBN have been studied by computer simulations of diamond, cBN, and hBN in Refs. [77, 78], of diamond in Refs. [1, 2, 79-86], of cBN and hBN in Refs. [52, 54, 86-89], of cBN in Refs. [55, 90-96], and of undoped hBN in Refs. [60-62, 68, 72, 76, 97-104] as there has been a significant progress in the theory of electronic states in solids [105] since Kohn-Sham's density functional theory in 1965 [106]. Many of the simulations used the supercell model to study both doped and undoped materials. The present paper was based on the Korringa-Kohn-Rostoker (KKR) scheme [107-110], and an open simulation code created by Akai [111, 112] was used. The feature of the simulation code will be described in the next section. Recent development of the KKR scheme and its fundamental advantage over the supercell model is described in detail by Ebert *et al.* [113], hence will not be described here. It should, however, be noted that the present work has been done in reference to many articles based on the supercell model.

In Sec. II, the computational procedure will be described. The computed results of the DOS and the band diagrams of diamond, cBN and hBN will be presented and discussed in Sec. III. Finally, conclusions will be given in Sec. IV.

II. Computational Procedure

The crystal structure of diamond is expressed by the space group $Fd3m$ ($O_h^7$) in which the primitive unit cell consists of two C atoms on two face-centered cubic (*fcc*) lattices; the second *fcc* lattice is displaced from the first one by ($a/4$, $a/4$, $a/4$) where $a$ is the lattice constant, $a = 0.3567$ nm [58]. There are vacant sites on two more *fcc* lattices displaced by ($2a/4$, $2a/4$, $2a/4$) and ($3a/4$. $3a/4$, $3a/4$) from the first one. The atomic C density of diamond is $1.77 \times 10^{23}$ /cm$^3$. The crystal structure of cBN is zinc-blende with the space group $F\text{-}43m$ ($T_d^2$). There are B and N atoms in the primitive unit cell, and the atomic density is $1.70 \times 10^{23}$ /cm$^3$. The lattice constant is $a = 0.36159$ nm [58]. There are vacant sites in the positions equivalent to diamond mentioned above. The crystal structure of hBN is expressed by the space group $P6_3/mmc$ ($D_{6h}^4$) with two B and two N atoms in the primitive unit cell. The lattice constants are $a = b = 0.25045$ nm and $c = 0.6606$ nm [58]. The atomic density is $1.15 \times 10^{23}$ /cm$^3$. The computed lattice constants of diamond and cBN at their energy minima were almost perfectly (within less than 1%) agreed with experimental values. In the present computations, the experimental lattice parameters of undoped materials were used for all doped materials like many of the previous works. Furthermore, no local lattice relaxation around the dopant atoms was taken into account. In particular, since hBN has a stacked layer structure, the substitution of B or N atoms with foreign atoms is likely to distort the planar structure in the real situation. Such a structural distortion was totally ignored like in the previous works [54, 55, 68, 93, 100]. The impurity atom doping was substitutional, and neither the interstitial case nor hydrogen impurity was considered. The dopant concentrations used were 0.0, 0.5, 1.0, 5.0, and 10.0 at.%, but other cases were computed, if necessary. As for the notations of doped materials, a *p*-type diamond with $x$ at.% B impurity atoms, for instance, will be denoted as *p*-D(C:$x$B). Similar notations will also be used for cBN and hBN in such a way as *n*-c(B:$x$Si)N and *p*-hB(N:$x$Be), respectively. The notations, *p*, *n*, and $x$, will be often omitted. For the case of c(B:$x$Be)(N:$y$Be), for instance, $x \neq y$ in general, but only the cases of $x = y$ were computed in the present work, and the same applies to doped hBN. It should also be noted that since the electronic bands are progressively broadened as the dopant concentration is increased, the terms of VBM and CBM will be used to express the peak positions of the highest valence band and the lowest conduction band, respectively.

To run the Akai-KKR code [112], the Linux g77 compiler (Fortran77) was used on desktop computers [114]. The basic theory and the manual of the code can be found in Refs. [112, 115], respectively. For the exchange-correlation potential parameters, those of Moruzzi, Janak, and Williams [116], built-in in the code, were used. The electron scattering computations in the code were set to be done up to *d*-orbitals, *i.e.*, for the angular quantum number $l \leq 2$. The number of *k*-points in the irreducible BZ for successful computations was 1814 for diamond and cBN, and 1376 for hBN. The energy mesh in the output was either 0.143 eV when the full energy range in computation was 2.0 Ry (27.2 eV), or 0.272 eV when it was 4.0 Ry (54.4 eV). Since the energy mesh was coarse, it is rational to depict the DOS diagram using a histogram [117], but it was depicted by a line graph for technical reasons. Since the code returns the band energies and the magnitude of Bloch spectral functions for specific reciprocal lattice vectors, the band diagrams were depicted with closed circles for the values of the Bloch spectral functions greater than a certain value that was determined so that the band diagram was best viewed. To present the band broadening, however, the band diagrams were depicted by histograms. The Akai-KKR code uses the Kohn-Sham theory [106] with the local density approximation (LDA) and the coherent potential approximation (CPA) [111, 113, 118]. The atomic-sphere approximation (ASA) [119] is optional. It is suggested to place empty spheres (ESs) in vacancies for better results [120-122]. The built-in CPA was useful as an arbitrary value was selected for the dopant concentration in the simulation [80, 112].

Some problems occurred in the computed results of hBN. In the first series of computations, the ASA was not used, but the CBM at the Γ-point was so low that hBN was a semiconductor of direct transition at the Γ-point, in contradiction to the results of the previous works [52, 59, 60, 68, 72, 98, 99]. In the second series of computations, the ASA was included, but the valence band structure was entirely different from those of the previous works. Most reasonable results were obtained in the third series of computations where the ASA was not used and two ESs were placed at the vacant sites between the B-N layers [87, 113]. Computations on diamond and cBN were also done in parallel to those on hBN, and the best results, in consistent with the previous works, were obtained by the use of the ASA and the ESs in the vacant sites. The ASA increased the indirect bandgap energy in diamond by 0.5 eV. The LDA is known to underestimate bandgap energies in semiconductors [95], and it was the case in the present computations. In summary, a KKR-LDA-CPA-ASA-ES scheme was used for diamond and cBN, while a KKR-LDA-CPA-ES scheme was used for hBN in the present paper. All approximations are built-in in the Akai-KKR code.

## III. Results and Discussion

A summary of the computed results is given in Table I. The notations, *i*, *p*, and *n* in the column of "Type", stand for intrinsic, *p*-type, and *n*-type semiconductors, respectively, and (*p*), for instance, indicates that the Femi level moved toward the valence band by doping but did not move below the VBM. The notation (*n*) has a similar meaning. The doped materials were degenerate in seven cases indicated by "degenerate" in the column of "Heavy doping". In case that localized states appeared like a band in the bandgap or a shoulder at the edge of the conduction or valence band in the DOS diagrams, it is notified in the column of "Localized states". In case that a doped material became degenerate by doping atom X, the dopant concentration, $n_X$, at which the Fermi level was located at the VBM or CBM, is noted in units of atomic % (at.%) in the column of "Fermi level at VBM or CBM". Finally, the column of "Bandgap" listed the bandgap energies computed in the present work as well as those in experiments. Note that the bandgap energy in the present work stands for the energy difference between the peaks of the CBM and the VBM in the computed band diagram, and not the one evaluated from the DOS diagram, because the energy difference was unambiguously determined in the former. Note also that in all figures throughout the present paper, the Fermi level is set to be the origin of the energy axis.

### A. Diamond

#### 1. Undoped diamond

The computed results of the DOS and the band diagram of undoped diamond are shown in Figs. 1(a) and 2, respectively. The feature of the DOS was similar to those of the previous works [1, 78, 80, 81, 86, 117, 122, 123]. The full width of the computed valence band was 21.8 eV, in reasonable agreement with an experimental width, 23.5 eV [19]. In the band diagram of Fig. 2, the indirect bandgap energy between the VBM at the Γ-point and the CBM at the X-point was 4.08 eV, 25% smaller than the experimental value, 5.47 eV [1]. The CBM was at the position approximately 3/4 from the Γ- to the X-points, in consistent with previous works [1, 2, 78, 86, 117, 123]. The Fermi level was located at 2.58 eV above the VBM. Figure 3 shows a bird-eye view of the band structure from the Γ- to the X-point depicted by a histogram of Bloch spectral functions. It is seen that every band is sharp and only slightly flared out at the bottom.

#### 2. B-doped diamond

In order to clearly see the feature of the electronic state, the DOS and the band diagram of $n_B$ = 10 at.%, the highest B concentration in the present work, are shown in Figs. 1(b) and 4, respectively. The overall shape of the DOS was similar to that of undoped diamond (see Fig. 1(a)) apart from the smooth feature and the shift toward the higher energy with respect to the Fermi level. It is seen in the band diagram of Fig. 4, the Fermi level was located -1.50 eV *below* the VBM at the Γ-point. The presence of the multiple points in each band is due to the band broadening. This is better seen in Fig. 5, where the band dispersion from the Γ- to the X-points is depicted by a histogram of the Bloch spectral functions. It is seen that virtually all bands are significantly broader than in the undoped case shown in Fig. 3. In Fig. 6, the Fermi level, measured from the VBM at the Γ-point, is plotted against $n_B$ that is equivalent to the hole concentration in theory. The data points refer to the ordinate on the left-hand side. The open circles are the data taken from simulations [80-84] and experiments [19, 29, 38, 124] using (111) oriented diamond, and the solid circles are the present results. Above $n_B \geq 0.5$ at.%, all data are closely located along a smooth concave curve, indicating that the result of the present simulation agree with those of other theories and experiments. In Fig. 6, the Fermi level was located at the VBM when $n_B$ = 0.3 at.% ($5.3 \times 10^{20}$ /cm$^3$). This value was comparable to the critical carrier concentration for the metal-insulator transition, $n_c = 4.5 \times 10^{20}$ /cm$^3$ (0.25 at.%) of Ref. [11] for (100) oriented B-doped films, $n_c = 2 \times 10^{20}$ /cm$^3$ (0.1 at.%) of Ref. [125] for polycrystalline B-doped films, and the critical carrier concentration of superconductor-to-insulator transition, $n_{si} = 4 \times 10^{20}$ /cm$^3$ (0.2 at.%) of Ref. [16] for (111), (100), and (110) oriented films. This indicates that the $n_B$-value at which the Fermi level is located at the VBM serves a good measure for superconductivity to occur at low temperature. Also shown in Fig. 6 by the solid curve, which refers to the ordinate on the right-hand side, is the experimental superconducting transition temperature $T_c$ after Ref. [16] as a function of the carrier concentration. Given the carrier concentration, the position of the Fermi level measured from the VBM, and the $T_c$ as well, can be identified from this figure.

As stated in Sec. II, no lattice relaxation due to impurity atom doping was considered in the present work. According to a 216-atom supercell calculation ($n_B$ = 0.46 at.%) [126], the B-C bond length was elongated approximately by 4%. A recent calculation using a 128-atom supercell ($n_B$ = 0.78 at.%) [80] resulted in a 2% elongation of the B-C bond and a 1.3% reduction of the neighboring C-C bonds, but such a change increased the Fermi level only by 25 meV, indicating that the effect of local lattice relaxation on the electronic state is quite small. Likewise, the increase in the lattice constant of B-doped diamond with $n_B$ = 4 at.% was only 0.3% from that of undoped diamond [127]. These results show that the lattice relaxation due to B-doping does not require a substantial change in the present results.

*3. N-doped diamond*

Shown in Fig. 7 is the energy difference between the Fermi level and the CBM as a function of the N concentration, $n_N$. The Fermi level, located at 1.5 eV below the CBM for undoped diamond, quickly approached the CBM, reached the CBM when $n_N$ = 0.4 at.%, and slowly moved above the CBM for $n_N$ > 0.4 at.%. This means that N atom is a good *n*-type dopant as long as it stays at the C site in diamond. Although it is widely believed that the substitutional N atoms create a donor level at 1.7 eV below the CBM due to the JT effect, the cause of the deep level and even the presence of the JT effect have not been unequivocally understood [42-44, 126]. Under these circumstances, the present result is expected to stimulate more detailed research of the electronic state in potentially *n*-type N-doped diamond.

B. cBN

*1. Undoped cBN*

The computed results of the DOS and the band diagram of undoped cBN are shown in Figs. 8(a) and 9, respectively. The Fermi level is set to be the origin of the energy scale for each case. These results were consistent with those of the previous works [52, 54, 55, 78, 86, 88, 89, 90-93, 95, 96, 104, 128]. The indirect bandgap energy between the VBM at the Γ-point and the CBM at the X points in Fig. 9 was 4.57 eV, 28% smaller than the experimental bandgap, 6.36 eV [50]. Note that the Fermi level was located at 4.36 eV above the VBM at the Γ-point, or only 0.21 eV below the CBM, as seen in Fig. 9, and these are the reference energy values in the following descriptions of doped cBN.

*2. Be-doped cBN*

Figures 8(b) - (d) show the DOS of c(B:10Be)N, cB(N:10Be), and c(B:5Be)(N:5Be), respectively. Each of the DOS for cB(N:10Be) and c(B:5Be)(N:5Be) had localized states in the bandgap due to Be 2*p*, while such a state was absent in c(B:10Be)N. The Fermi levels in cB(N:10Be) and c(B:5Be)(N:5Be) were positioned at the localized states as if they were pinned by the states.

Figure 10(a) shows the Fermi level in c(B:Be)N measured from the VBM *vs.* the Be concentration, $n_{Be}$. The Fermi level moved down toward the VBM fairly quickly with $n_{Be}$, indicating that c(B:Be)N is a *p*-type semiconductor. It was then located on the VBM when $n_{Be}$ = 0.9 at.% (1.53 × 10$^{21}$ /cm$^3$), and moved below the VBM for $n_{Be}$ > 0.9 at.%, although the downward shift slowed down as $n_{Be}$ was increased. Experimentally, Be-doped cBN is shown to be *p*-type [49, 129], and since the substitution of B with Be has a low energy of formation [55, 130], it is likely that the B sites are preferentially substituted with Be atoms in experiment [131]. This implies that a degenerate $p$-c(B:Be)N, that exhibits a metallic conduction, is realized, if c(B:$x$Be)N with $x \geq 0.9$ at.% is synthesized. For the case of cB(N:10Be), the Fermi level was located at around 1.6 eV above the VBM at the Γ-point for 0.5 at.% $\leq n_{Be} \leq$ 10 at.% (the data not shown, but see Fig. 8(c)). The situation was similar for c(B:Be)(N:Be). The Fermi level initially shifted toward the valence band by more than 3 eV by Be-doping, but stayed roughly in the same position for higher $n_{Be}$ (see Fig. 8(d)) presumably due to the presence of the localized states. In Ref. [97], the donor and acceptor levels due to doping of various impurity atoms are computed using a 432-atom supercell, and c(B:Be)N with $n_{Be} = 0.23$ at.% is confirmed to be a $p$-type semiconductor, consistent with the present work. The computed acceptor level is as shallow as 0.16 eV, implying that the degenerate situation occurs at fairly low $n_{Be}$. On the other hand, in Ref. [130], it is mentioned that cB(N:Be) "leads to a deep mid-gap level". All these results are consistent with the present results of Figs. 8(c) and 10(a).

*3. Si-doped cBN*

Figures 11(b) - (d) show the DOS of c(B:10Si)N, cB(N:10Si), and c(B:5Si)(N:5Si), respectively. The doping of Si substituting B, *i.e.*, c(B:Si)N, is assumed to be $n$-type as Si atoms release electrons. In fact, the Fermi level moved toward the conduction band by Si-doping, as seen in Fig. 11(b); the Fermi level of c(B:10Si)N was located above the CBM. On the other hand, in cB(N:10Si), there was a shoulder due to Si 3$p$ at the edge of the valence band. The Fermi level shifted toward the valence band by roughly 3 eV from the position of undoped cBN, indicating only a weak $p$-type nature of cB(N:Si), but was away from the VBM for higher Si concentrations. For the case of c(B:5Si)(N:5Si), there was also a shoulder due to Si 3$p$ at the edge of the valence band. The behavior of the Fermi level with respect to $n_{Si}$ was complex, but the Fermi level always stayed in the bandgap.

The Fermi level of c(B:Si)N, measured from the CBM at the X-point is plotted against $n_{Si}$ in Fig. 10(d); it first moved toward the CBM from the undoped position with $n_{Si}$, crossed the CBM at $n_{Si} = 0.3$ at.% ($5.1 \times 10^{20}$ /cm$^3$), and moved further into the conduction band with $n_{Si}$. The present results indicated that only c(B:Si)N is a good $n$-type semiconductor. This supports the experimental result that a Si-doped cBN is $n$-type [49, 56, 57]; however, judging from the larger atomic size of Si than B and N atoms, it is likely that Si atoms also create a large change in the local structure no matter where they are situated, and the local distortion and dangling bonds around Si atoms also make Si-doped cBN $n$-type.

In a simulation work, a donor level is created in c(B:Si)N [97], indicating that it is $n$-type,

in agreement with the present work. In Refs. [55, 93], and the Fermi level of undoped cBN is located at the edge of the valence band, unlike the present result shown in Fig. 11(a), and for the case of c(B:1.56Si)N, there is a sharp band in the center of the bandgap, and the Fermi level is pinned at the position. Such a band was absent in all c(B:Si)N cases in the present work. Regarding cB(N:Si) in the present work, there were weak localized states overlapping the edge of the valence band, and the Fermi level was pinned at the position (e.g. see Fig. 11(d)). In Ref. [53], the resistivity of Si-doped cBN films drops to approximately 1/10 from that of the undoped film when $n_{Si}$ = 1.5 at.%, but is still too high to undertake Hall measurements. This is consistent with the present result of c(B:Si)(N:Si) that the Fermi level did not move closer to the VBM by Si-doping up to $n_{Si}$ = 10 at.% (see Fig. 11(d)) after the initial downward shift in $n_{Si}$ ≤ 0.5 at.%.

*4. C-doped cBN*

Figures 12(b) - (d) show the DOS diagrams of c(B:10C)N, cB(N:10C), and c(B:5C)(N:5C), respectively. The doping of C substituting B, *i.e.*, c(B:C)N, can be assumed to be *n*-type as C atoms release electrons. Indeed, the Fermi level moved toward the conduction band by C-doping, as seen in Fig. 12(b). By contrast, the Fermi level of cB(N:10C) moved to the edge of the valence band, showing that it is *p*-type. For the case of c(B:5C)(N:5C), the Fermi level was located nearly in the center of the bandgap. It appears as if the *n*-type nature of c(B:C)N and the *p*-type nature of cB(N:C) were canceled out, and indeed, it was confirmed in the present work of c(B:$x$C)(N:$y$C), that the Fermi level was shifted, depending on the $x$ and $y$ values.

Shown in Fig. 10(b) and (c) are the shifts of the Fermi levels in *n*-c(B:C)N and *p*-cB(N:C) *vs.* the C concentration, $n_C$. The Fermi level of *n*-c(B:C)N, measured from the CBM at the X-point, moved toward the CBM as $n_C$ was increased, crossed the CBM at $n_C$ = 0.3 at.% (0.5 × $10^{21}$ /cm$^3$), and moved above the CBM as $n_C$ was increased further. By contrast, the Fermi level of *p*-cB(N:C), measured from the VBM at the Γ-point, quickly moved down to the valence band as $n_C$ was increased, crossed the VBM at $n_C$ = 0.9 at.% (1.5 × $10^{21}$ /cm$^3$), and gradually moved down into the valence band. To sum up, the position of the Fermi level for both *n*-c(B:C)N and *p*-cB(N:C) sensitively depended on the C-doping concentration, and both doped semiconductors were degenerate for heavy C-doping.

C. hBN

*1. Undoped hBN*

The computed DOS and the band diagram of undoped hBN are shown in Figs. 13(a) and 14, respectively. The main feature of the DOS diagram was consistent with the previous results [54, 60-62, 99]. In the band diagram of Fig. 14, the valence band facing to the bandgap was fairly flat between the A-Γ and M-L points, and the conduction band was the lowest at the Γ-point. The indirect bandgap energy was the smallest between the M-L points in the valence band and the Γ-point in the conduction band, and the bandgap energy was 3.7 eV, only 57 - 60% of the experimental values, 6.4 eV [70], 6.5 eV [69] and 6.17 eV [51]. Apart from the fact that there were flat portions in the upper part of the valence band facing to the bandgap, the details of the computed band diagram were significantly different from previous reports [52, 60, 62, 68, 72, 78, 86, 89, 96, 98, 99, 100, 103, 104]; most noticeable was the low CBM at the Γ-point. The reason for this has not been identified. In Ref. [99], the bandgap minimum is between the VBM at the H-point and the CBM at the M-point. Many of the previous works obtained similar results. In Ref. [62], however, the Γ-point in the conduction band is the closest in energy to both the M-point and the flat band between the K- and H-points in the valence band, similar to the present result. In Ref. [99], the concave conduction band at the Γ-point is assigned to the interlayer band between B-N hexagonal planes. If this is actually the case, it follows that the interlayer electronic interaction was stronger in the present simulation than in other works.

*2. Be-doped hBN*

The DOS diagrams of h(B:10Be)N, hB(N:10Be), and h(B:5Be)(N:5Be) are shown in Figs. 13(b) - (d), respectively. The Fermi level of h(B:10Be)N was located at the upper edge of the valence band so that it was a *p*-type semiconductor [97]. There was no localized state in the bandgap. According to the band diagram (not shown) of h(B:Be)N with $n_{Be} \geq 2$ at.%, the Fermi level was approximately 0.05 eV below the VBM in the following regions: (i) the Γ-point, (ii) from the A- to the Γ-point, and (iii) from the M-point to the half-way to the L-point. It was thus predicted that h(B:Be)N could be metallic in the $n_{Be}$ range. It is therefore of great interest to study whether h(B:Be)N becomes superconducting because unlike B-doped diamond, there are multiple points in the valence band where the Fermi level is below the valence band. If hB(N:Be) is coexistent in experiment, however, the Fermi level is likely to be pinned by at the localized states, and h(B:Be)N is no longer *p*-type or metallic, as inferred from Fig. 13(d). On the other hand, for the case of hB(N:10Be), there were localized states due to Be 2*p* in the middle of the bandgap, as seen in Fig. 13(c), and the Fermi level was pinned at the same position. This was surprising because hB(N:10Be) is expected to be *p*-type, and the Fermi level should be close to the valence band. Similarly, for the case of

h(B:5Be)(N:5Be), there were localized states due to Be 2*p* in the middle of the bandgap, as seen in Fig. 13(d), and the Fermi level was pinned at the same position.

*3. C-doped hBN*

The DOS diagrams of h(B:10C)N, hB(N:10C), and h(B:5C)(N:5C) are shown in Figs. 15(b) - (d), respectively. It is seen that none of the cases is degenerate. The Fermi level in h(B:C)N shifted up by approximately 1 eV toward the conduction band by C-doping with $n_C$ up to 1 at.%, and stayed approximately in the same position for $n_C > 1$ at.% (data not shown). In Fig. 15(b), there were localized states due to C 3*s* near the conduction band, and the Fermi level was pinned at the position. The Fermi level in hB(N:C) moved down only by 0.8 eV from the undoped position to the valence band by C-doping. The Fermi level was located in the same position as the localized states due to C 3*s* in the bandgap. For the case of h(B:C)(N:C), the shift of the Fermi level by C-doping was also small, but as seen in Fig. 15(d), the Fermi level was not located at the localized states due to C 3*s* at ~ 1.5 eV but at the edge of the valence band. It should be noted that hB(N:C) is actually synthesized [140], and C atoms are present in the atomic positions of hBN [135]. In Ref. [97], it is concluded that h(B:C)N and hB(N:C) are *n*- and *p*-type, respectively, but the impurity levels are so deep ( >1 eV) that substitutional dopants are not effective for semiconducting hBN. This result is consistent with the present results in the sense that the shift of the Fermi level by C-doping was small even though $n_C$ was increased to as high as 10 at.%.

IV. Conclusion

The electronic states of doped diamond, cBN, and hBN were computed using the KKR scheme to investigate the shift of the Fermi level by impurity atom doping, which took an advantage of the built-in CPA in the Akai-KKR code used. The dopant atoms were B and N for diamond, Be, Si, and C for cBN, and Be and C for hBN, and the computations were done for the dopant concentration up to 10 at.%. It was found that the Fermi level was located on the VBM when $n_B = 0.3$ at.% for B-doped diamond, when $n_{Be} = 0.9$ at.% for c(B:Be)N, when $n_C = 0.9$ at.% for cB(N:C), and when $n_{Be} \sim 2$ at.% for h(B:Be)N. On the other hand, the Fermi level was located on the CBM when $n_N = 0.4$ at.% for N-doped diamond, if N atoms stayed at the diamond lattice sites, when $n_{Si} = 0.3$ at.% for c(B:Si)N, and when $n_C = 0.3$ at.% for c(B:C)N. The Fermi level of c(B:*x*C)(N:*y*C) was variable, depending on the *x* and *y*-values. These $n_X$ values are the critical concentrations for semiconductor-to-metal transition, and in particular, serve a measure for the occurrence of superconductivity at low temperature for B-doped diamond. It was also found that the degeneracy occurred only in the seven cases in which there was no localized state in the bandgap or at the band edge. The possibility that the degenerate h(B:Be)N could be a unique superconductor was suggested. Better approximations for exchange-correlation between electrons and the incorporation of van der Waals force in the KKR scheme are important in studying covalently-bonded semiconductors consisting of light elements.

Acknowledgement

   The author thanks Hisazumi Akai and Kenji Watanabe for useful discussion, Osamu Mishima for supplying useful documents, Etienne Gheeraert, Paulina Plochocka, and Hisao Kanda for assisting the paper submission, and Jeffrey T. Glass for kind encouragement. The author is indebted to Hiroshi Kawarada for permission to use the data in Ref. [16]. Finally, the author is most thankful to Hitoshi Gomi for critical comments and useful discussion that made the author revise the present paper.

Table 1. Summary of computed results on diamond, cBN, and hBN. For the notations, see Sec. 3 in the text.

| Materials | Type | Heavy doping | Localized states | | Fermi level at VBM or CBM | Bandgap (eV) | |
|---|---|---|---|---|---|---|---|
| | | | Bandgap | Band edge | $n_X$ (at.%) | This work | Exp. |
| Diamond | $i$ | - | - | - | - | 4.08 | 5.47 [1] |
| D(C:B) | $p$ | degenerate | - | - | 0.3 | | |
| D(C:N) | ($n$) | - | - | - | - | | |
| Cubic BN | $i$ | - | - | - | - | 6.75 | 6.36 [50] |
| c(B:Be)N | $p$ | degenerate | - | - | 0.9 | | |
| cB(N:Be) | ($p$) | - | present | - | - | | |
| c(B:Be)(N:Be) | ($p$) | - | present | - | - | | |
| c(B:Si)N | ($n$) | degenerate | - | - | 0.3 | | |
| cB(N:Si) | ($p$) | - | - | present | - | | |
| c(B:Si)(N:Si) | ($n$) | - | - | present | - | | |
| c(B:C)N | $n$ | degenerate | - | - | 0.3 | | |
| cB(N:C) | $p$ | degenerate | - | - | 0.9 | | |
| c(B:C)(N:C) | $p, n$ | degenerate | - | - | - | | |
| Hexagonal BN | $i$ | - | - | - | - | 3.7 | 5.7 [66, 68] |
| h(B:Be)N | $p$ | degenerate | - | - | ~2 | | |
| hB(N:Be) | - | - | present | - | - | | |
| h(B:Be)(N:Be) | - | - | present | - | - | | |
| h(B:C)N | ($n$) | - | present | - | - | | |
| hB(N:C) | ($p$) | - | - | present | - | | |
| h(B:C)(N:C) | - | - | present | - | - | | |

Figure Captions

FIG. 1. DOS of (a) undoped and (b) B-doped diamond ($n_B$ = 10 at.%). The origin of the energy is the Fermi level.

FIG. 2. Band diagram of undoped diamond. The origin of the energy is the Fermi level.

FIG. 3. Bloch spectral functions of undoped diamond from the Γ- to the X-points, *i.e.*, along the (100) direction.

FIG. 4. Band diagram of B-doped diamond with $n_B$ = 10 at.%.

FIG. 5. Bloch spectral functions of B-doped diamond from the Γ- to the X-points, *i.e.*, along the (100) direction, for $n_B$ = 10 at.%.

FIG. 6. Fermi level measured from the VBM at the Γ-point *vs.* $n_B$ (at.%) in diamond: the open circles are the results of Refs. [19, 29, 38, 80, 82 - 84, 124], and the closed circles are the present results. The solid curve, which refers to the ordinate on the right-hand side, is the superconducting transition temperature $T_c$ after Ref. [16] (by courtesy of H. Kawarada).

FIG. 7. The Fermi level measured from the CBM at the X-point *vs.* $n_N$ (at.%) in N-doped diamond. The Fermi level moves up as the data points become smaller.

FIG. 8. DOS of undoped and Be-doped cBN: (a) undoped, (b) c(B:10Be)N, (c) cB(N:10Be), and (d) c(B:5Be)(N:5Be).

FIG. 9. Band diagram of undoped cBN.

FIG. 10. Fermi level in (a) *p*-c(B:Be)N measured from the VBM *vs.* the Be concentration, $n_{Be}$ (at.%), (b) *p*-cB(N:C) measured from the VBM *vs.* the C concentration $n_C$ (at.%), (c) *n*-c(B:C)N measured from the VBM *vs.* the C concentration, $n_C$ (at.%), and (d) *n*-c(B:Si)N measured from the CBM *vs.* the Si concentration, $n_{Si}$ (at.%).

FIG. 11. DOS of undoped and Si-doped cBN: (a) undoped, (b) c(B:10Si)N, (c) cB(N:10Si),

and (d) c(B:5Si)(N:5Si).

FIG. 12. DOS of undoped and C-doped cBN: (a) undoped, (b) c(B:10C)N, (c) cB(N:C), and (d) c(B:5C)(N:5C).

FIG. 13. DOS of undoped and Be-doped hBN: (a) undoped, (b) h(B:10Be)N, (c) hB(N:10Be), and (d) h(B:5Be)(N:5Be).

FIG. 14. Band diagram of undoped hBN.

FIG. 15. DOS of undoped and C-doped hBN: (a) undoped, (b) h(B:10C)N, (c) hB(N:10C), and (d) h(B:5C)(N:5C).

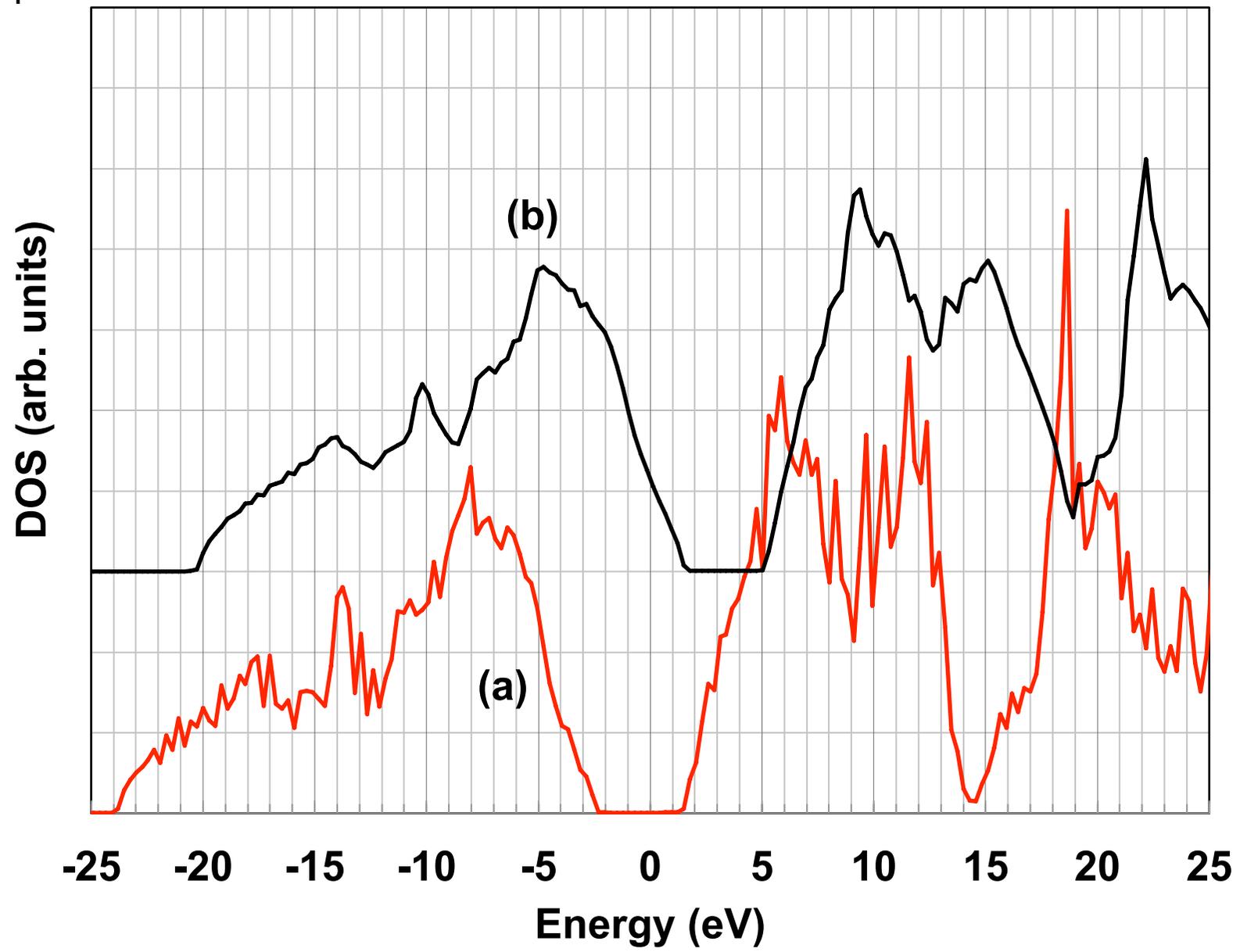

Fig. 1

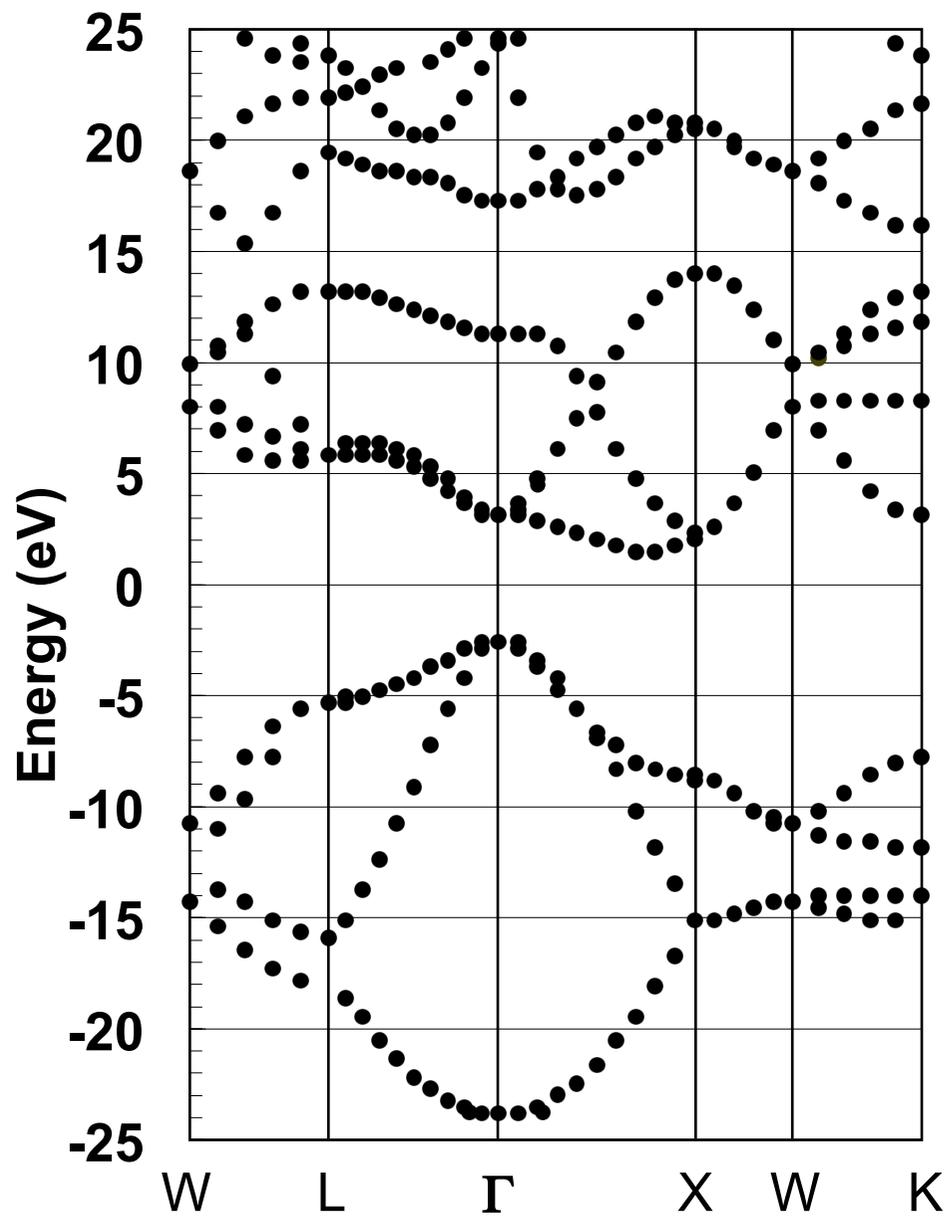

Fig. 2

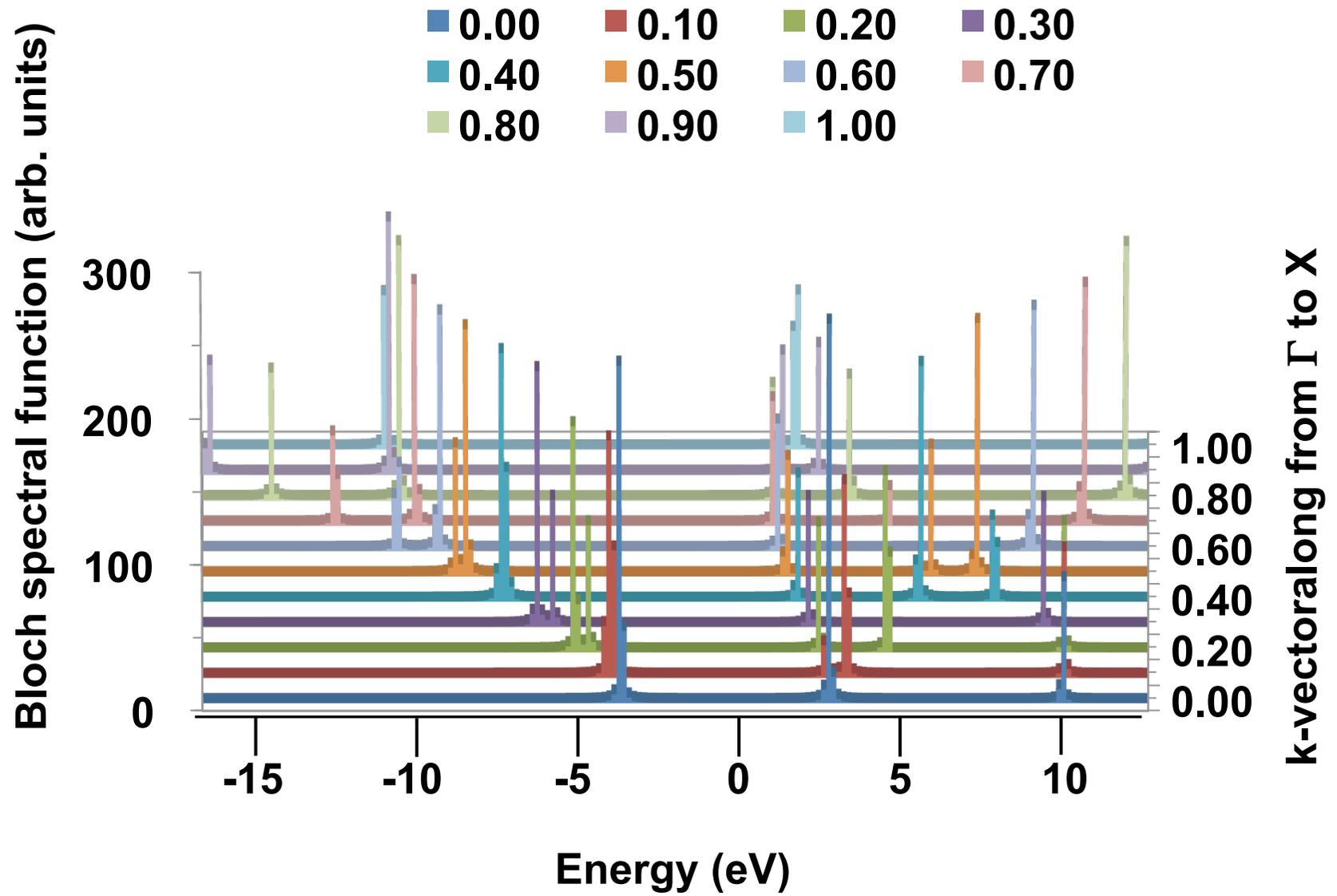

Fig. 3

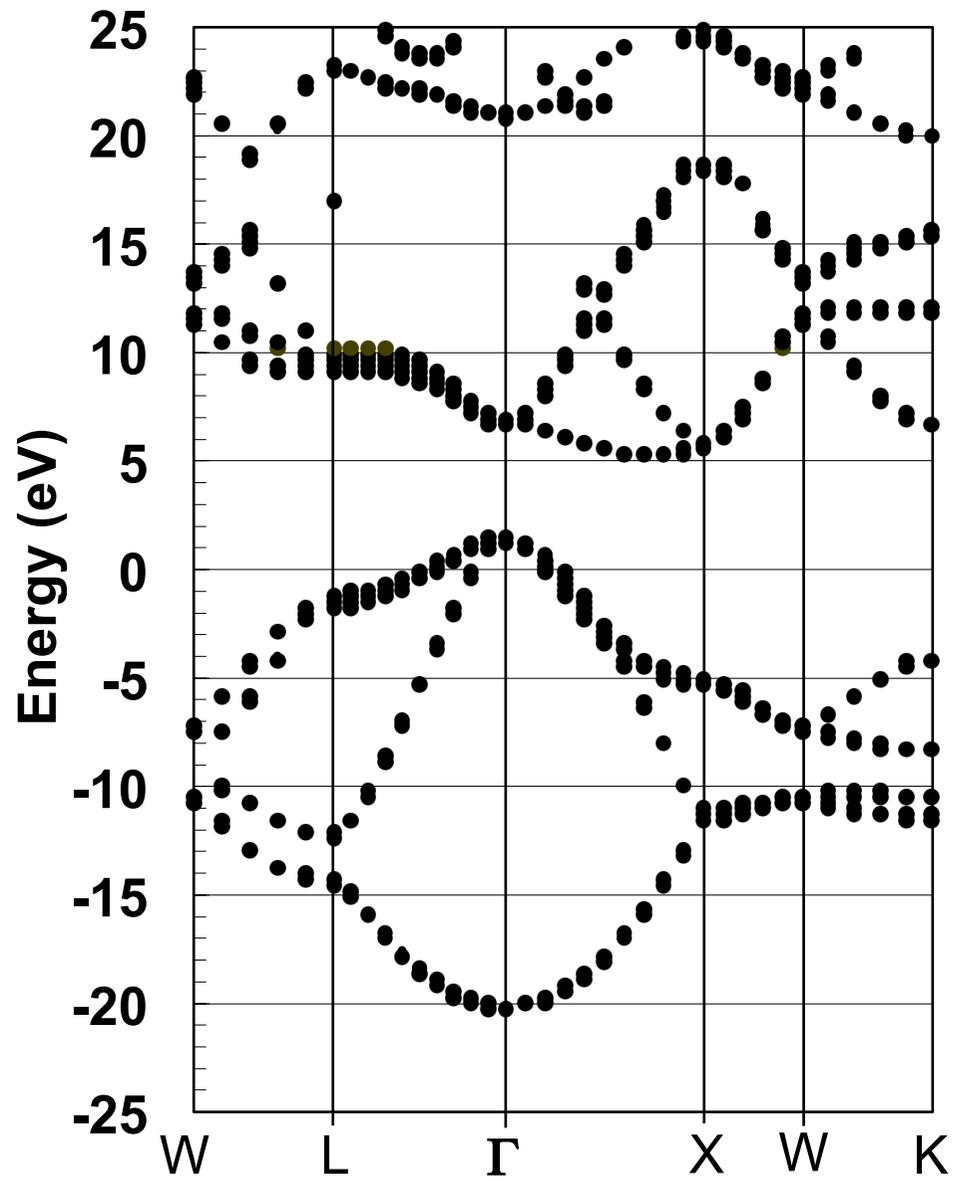

Fig. 4

Fig. 5

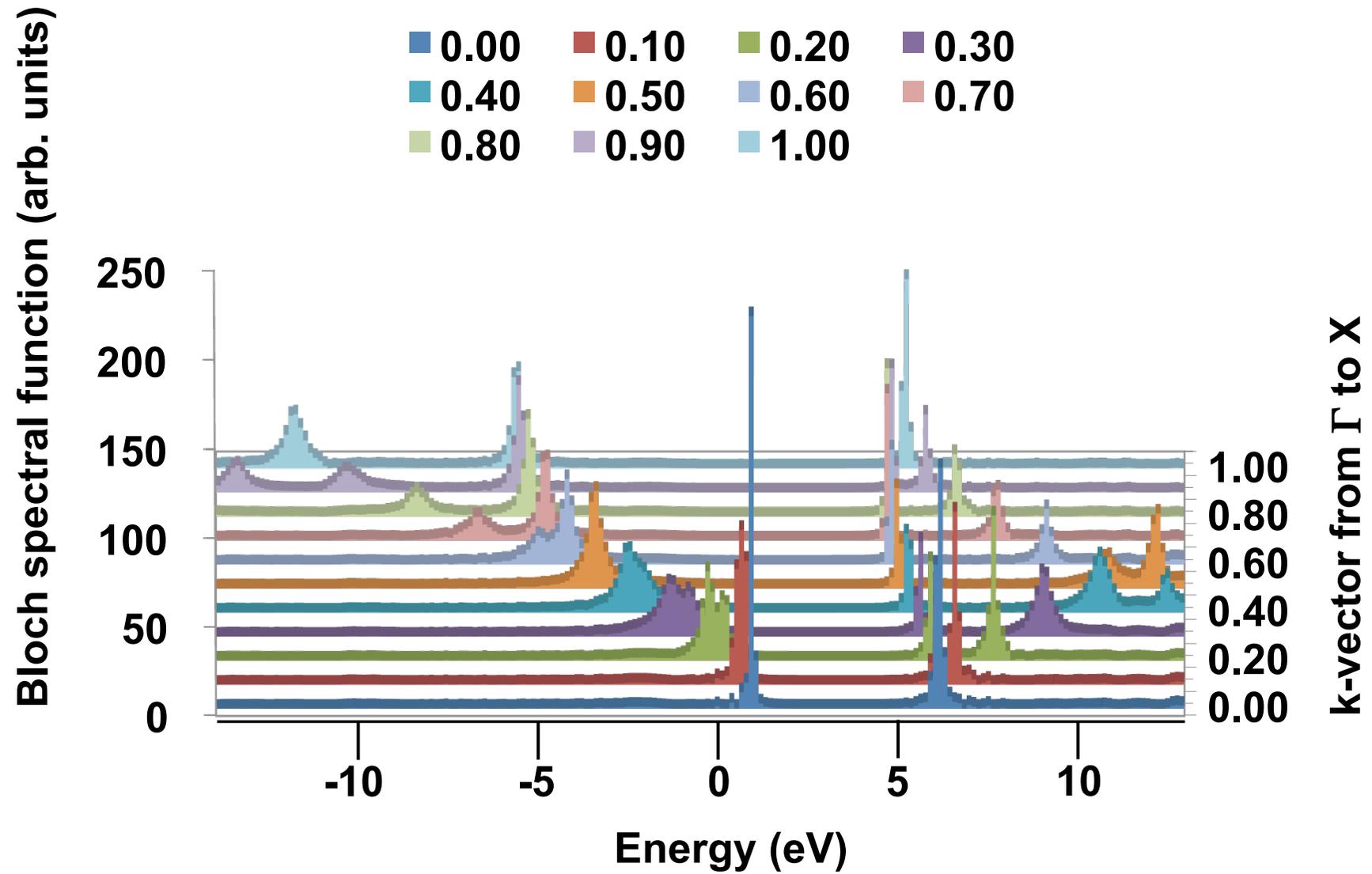

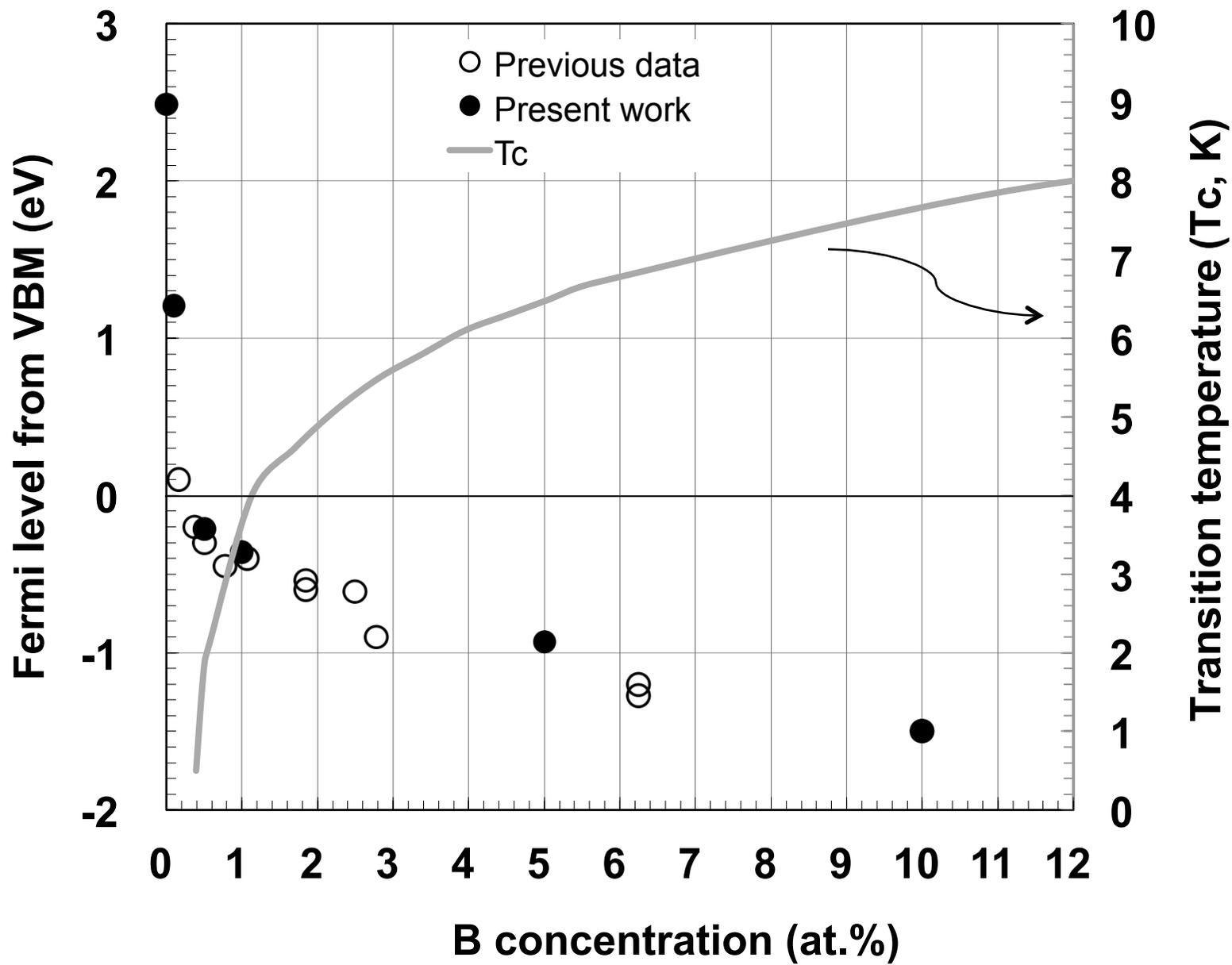

Fig. 6

Fig. 7

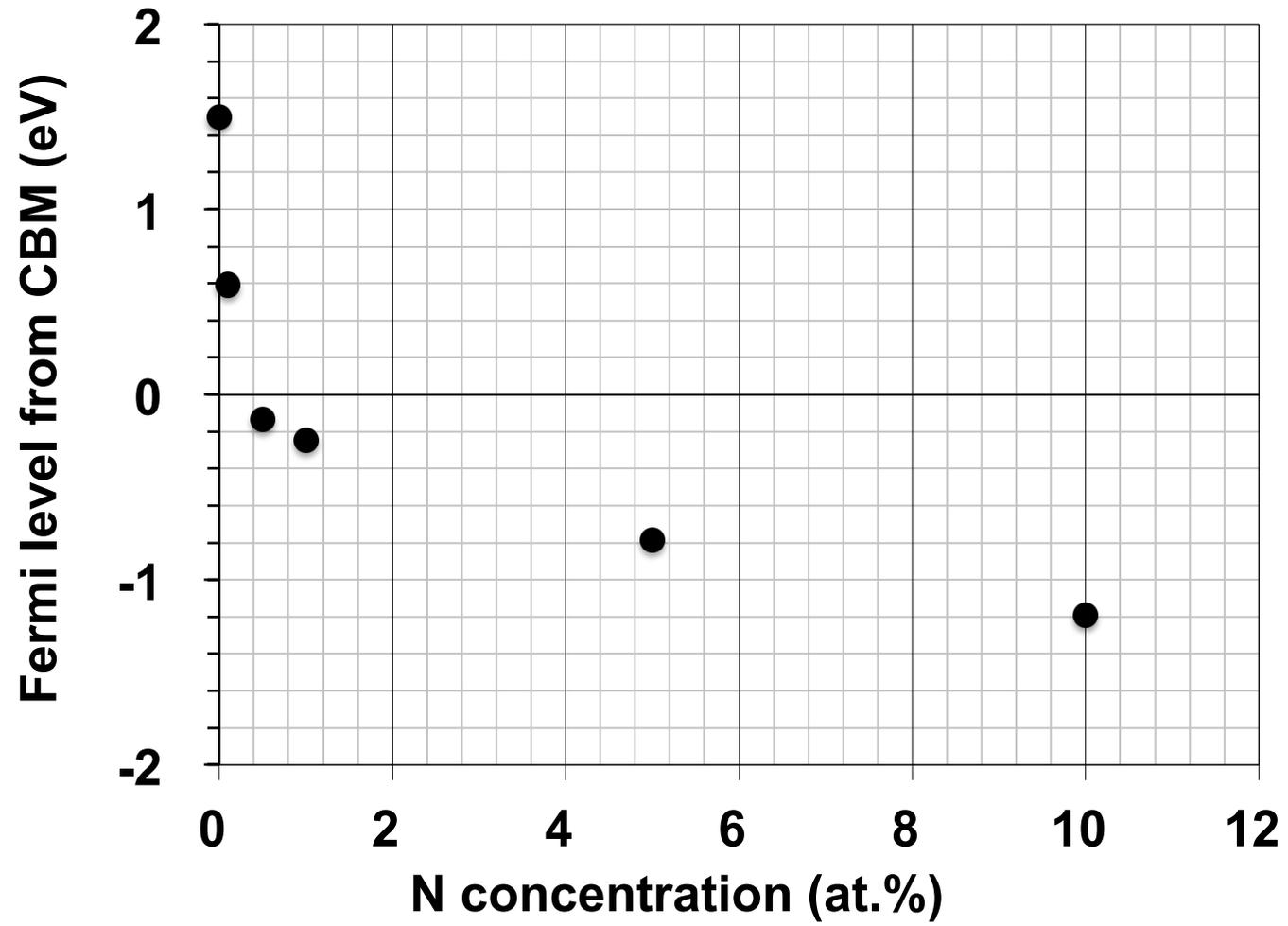

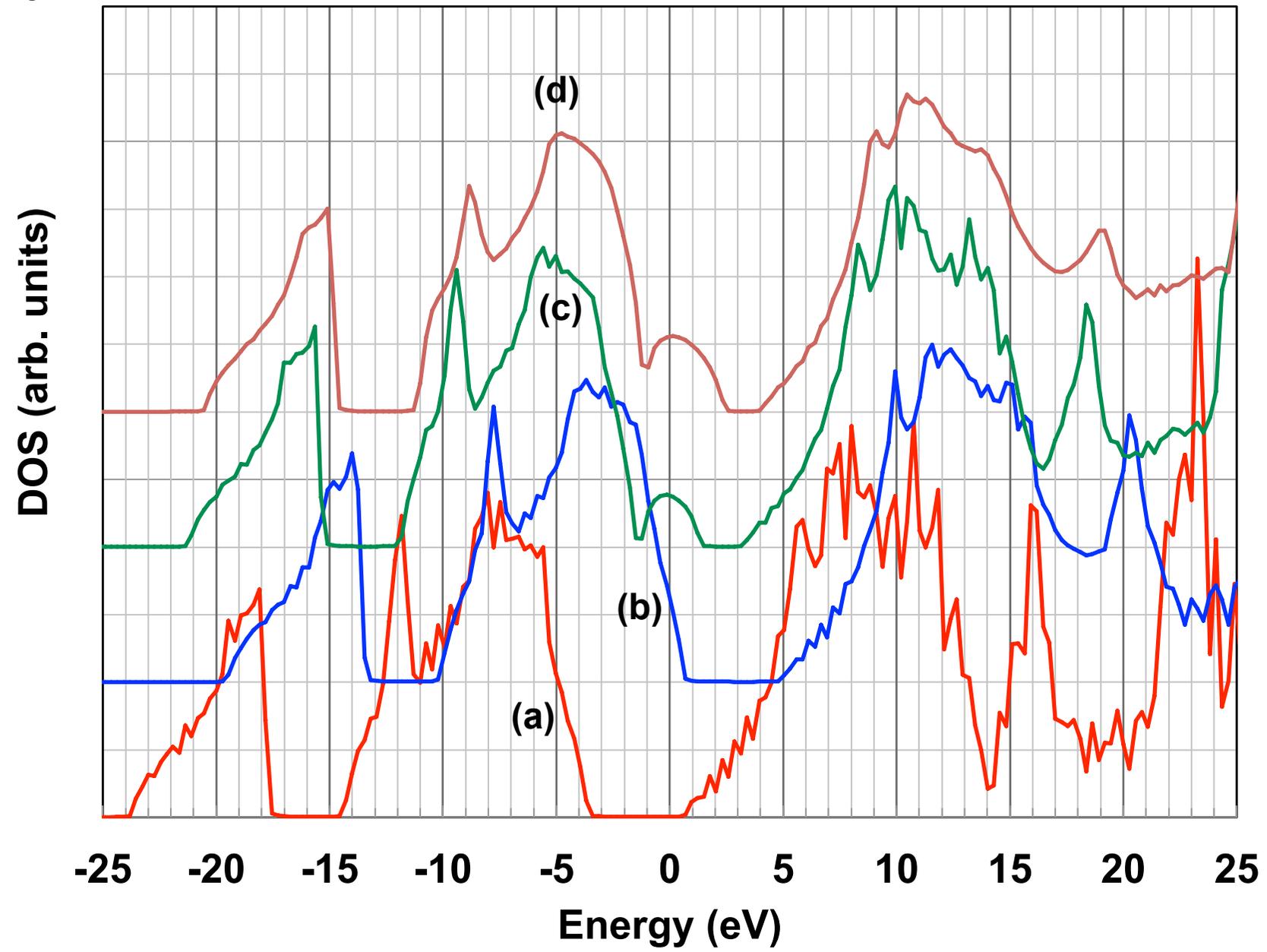
Fig. 8

Fig. 9

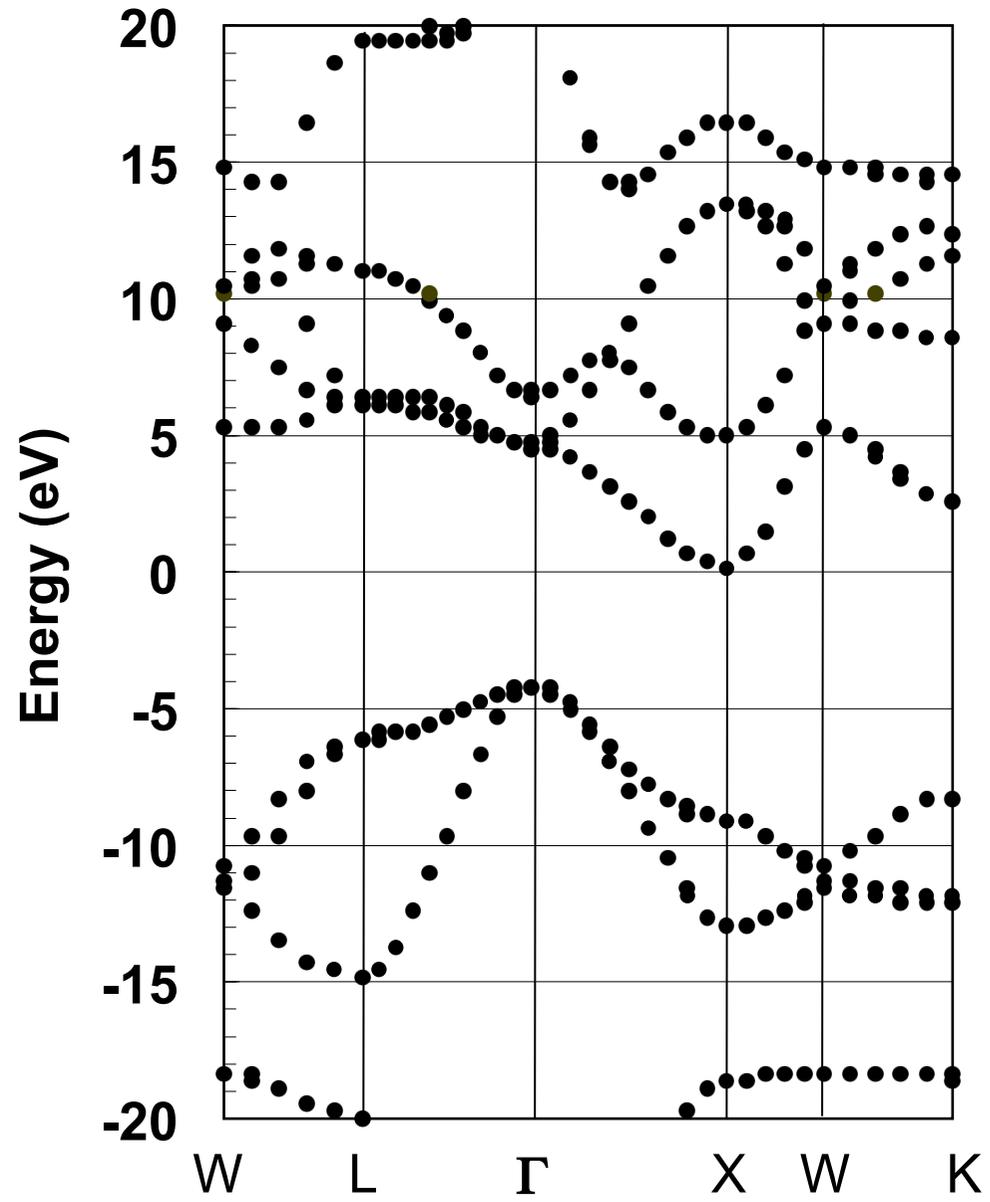

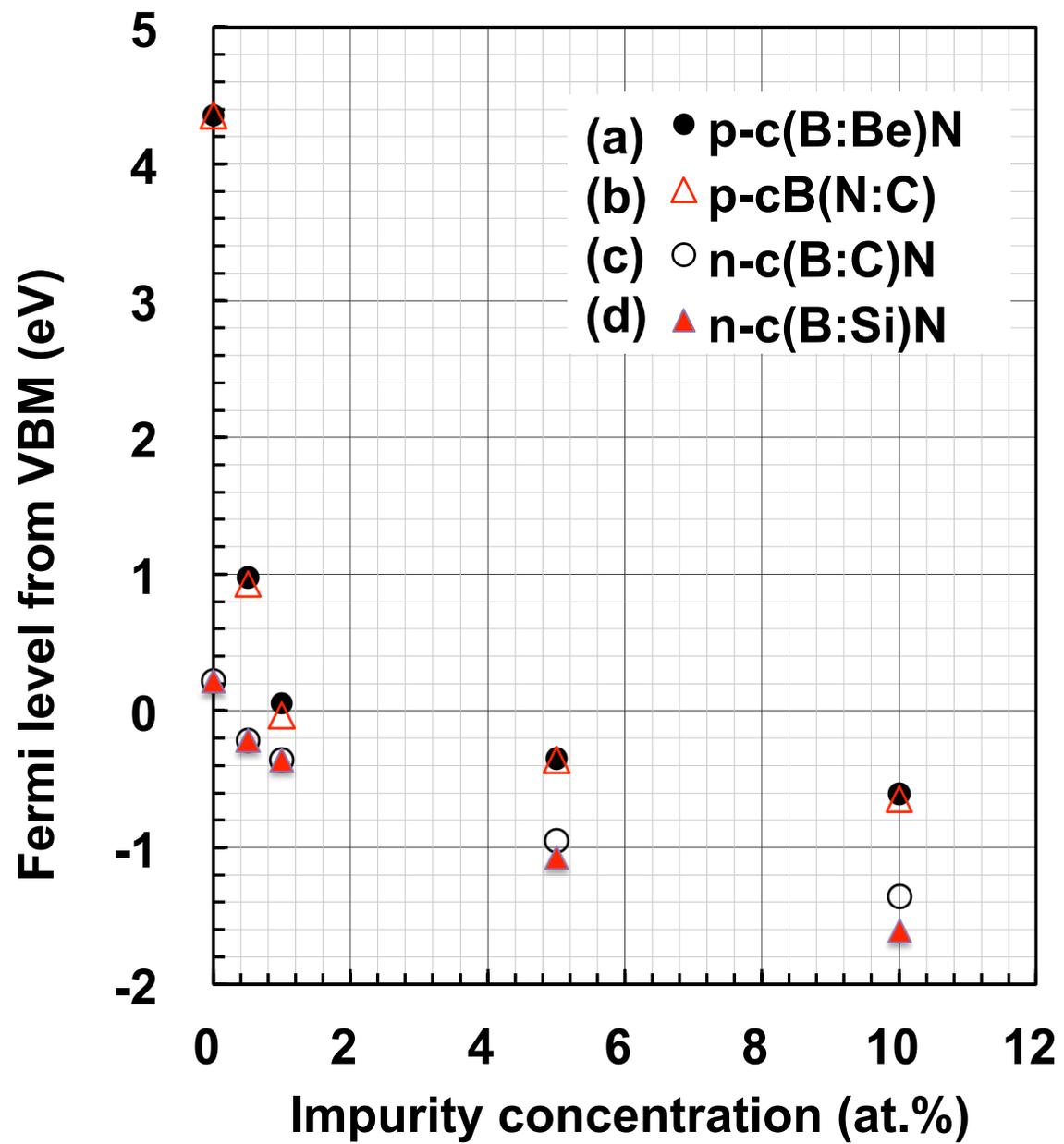

Fig. 10

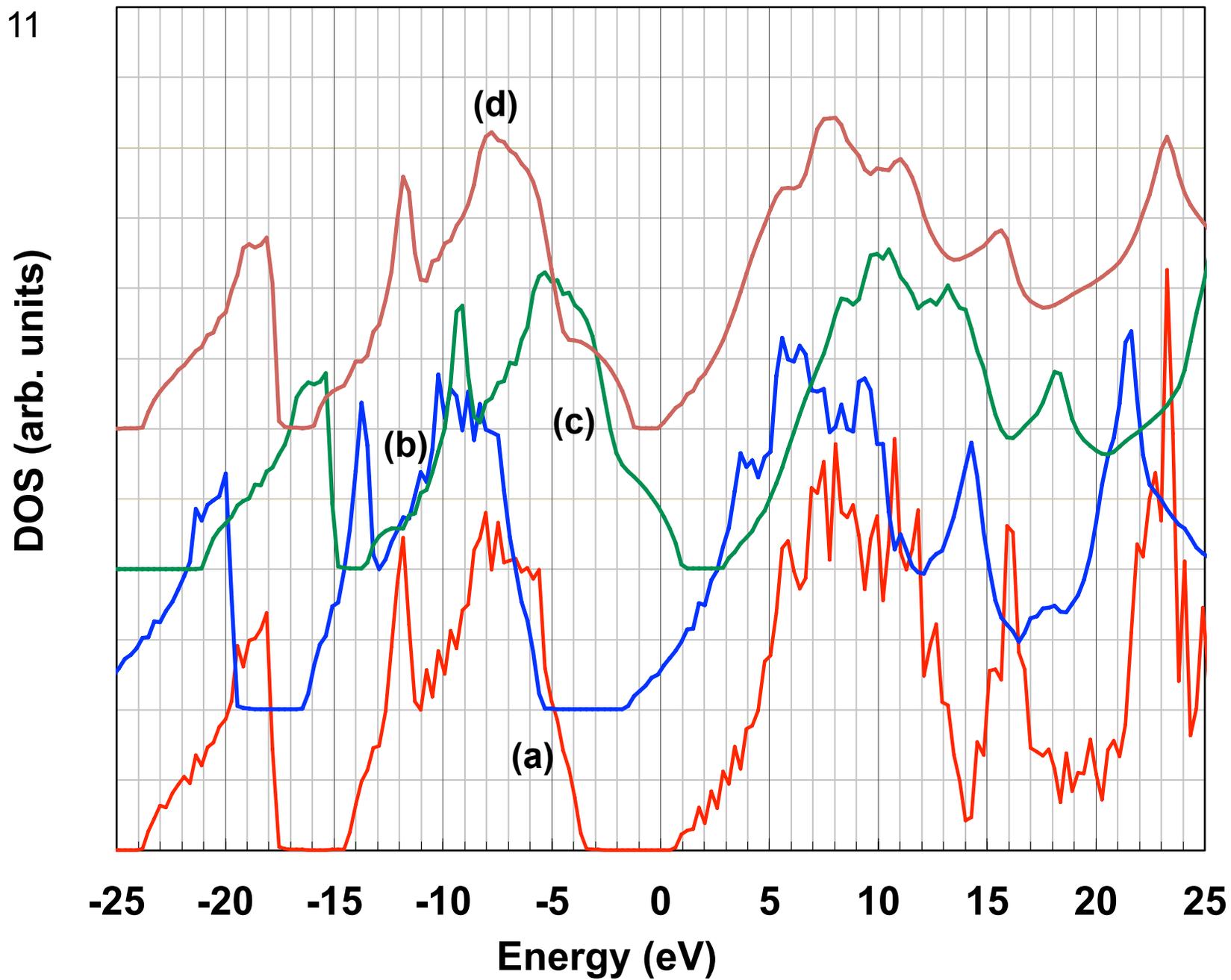

Fig. 11

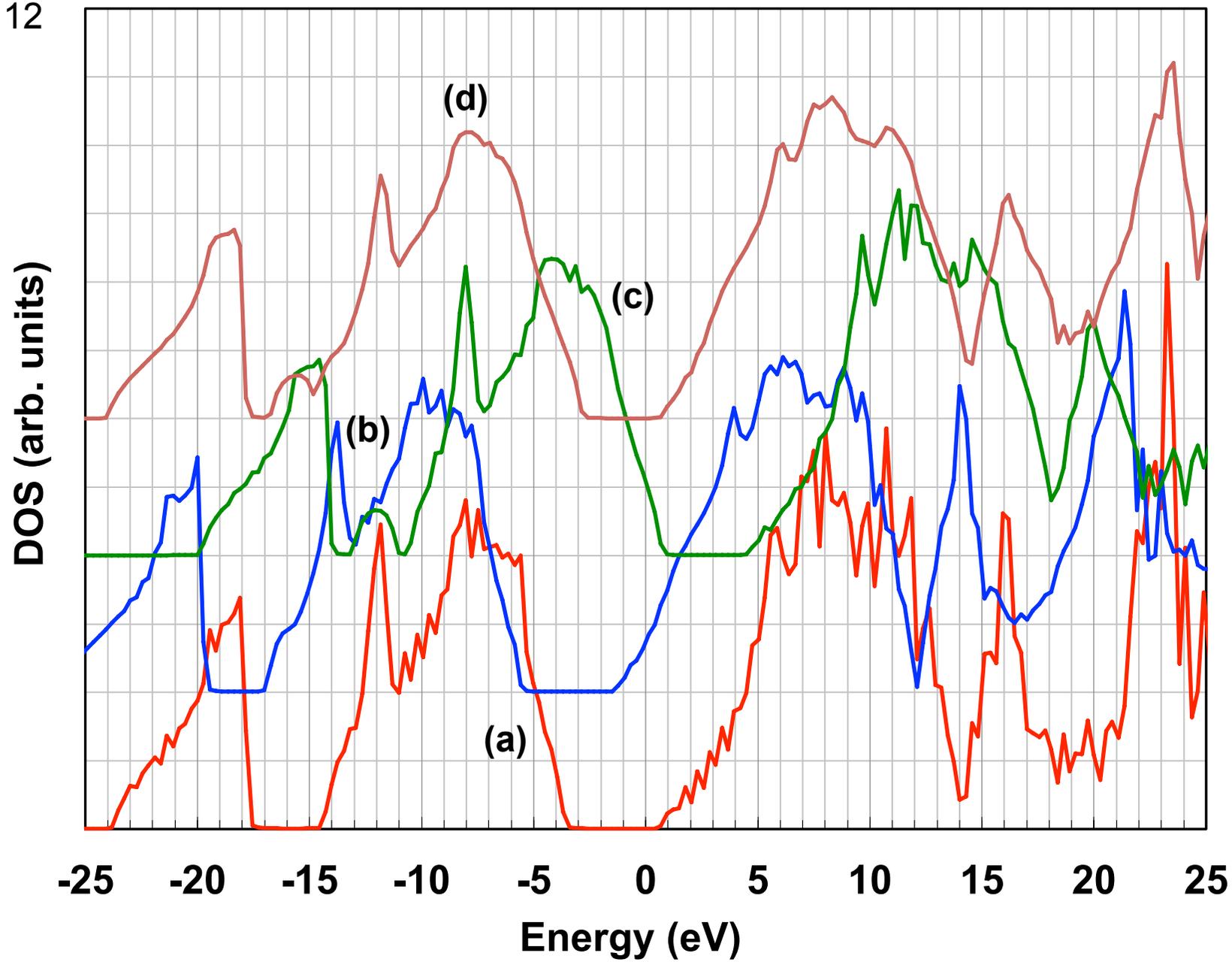

Fig. 12

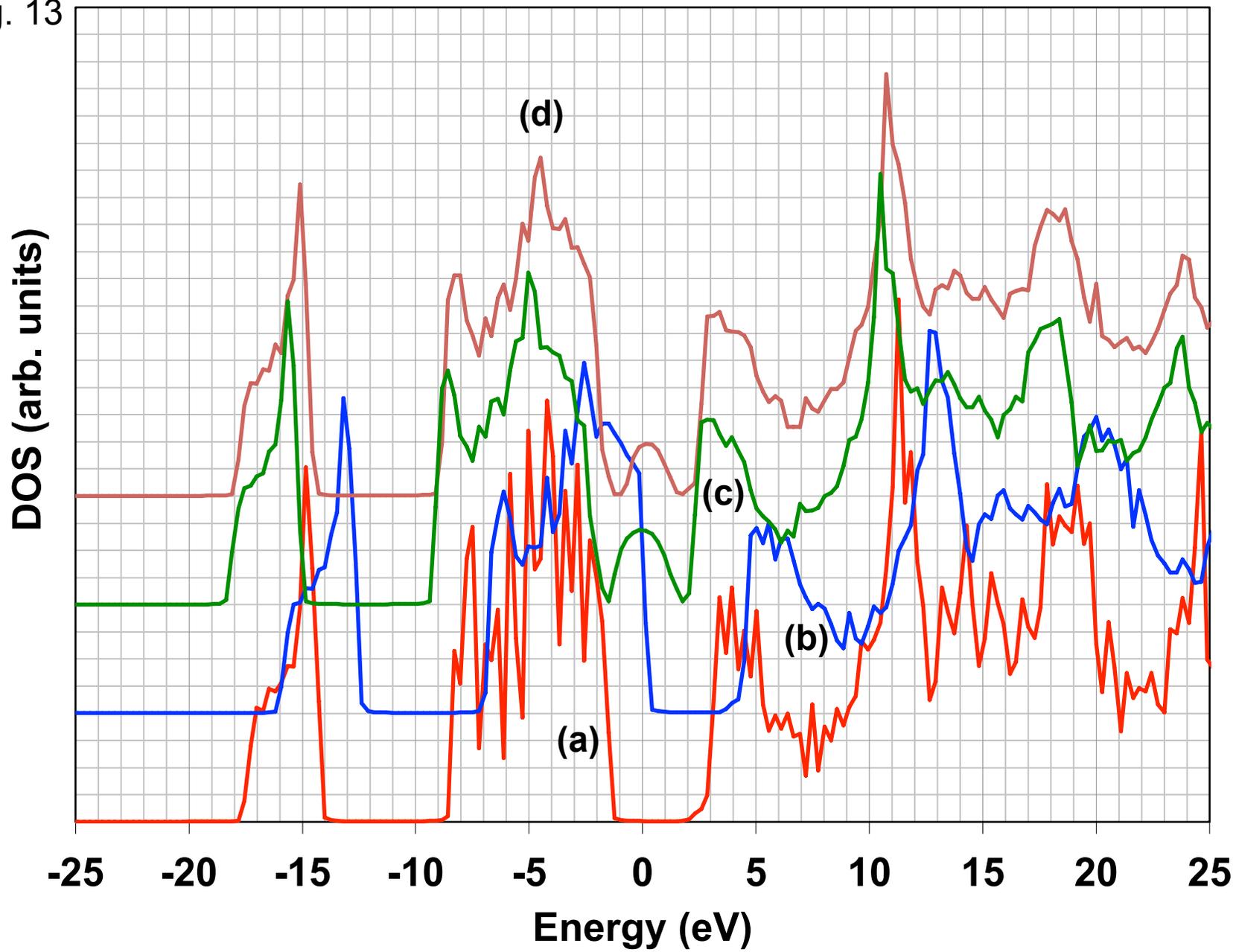

Fig. 13

Fig. 14

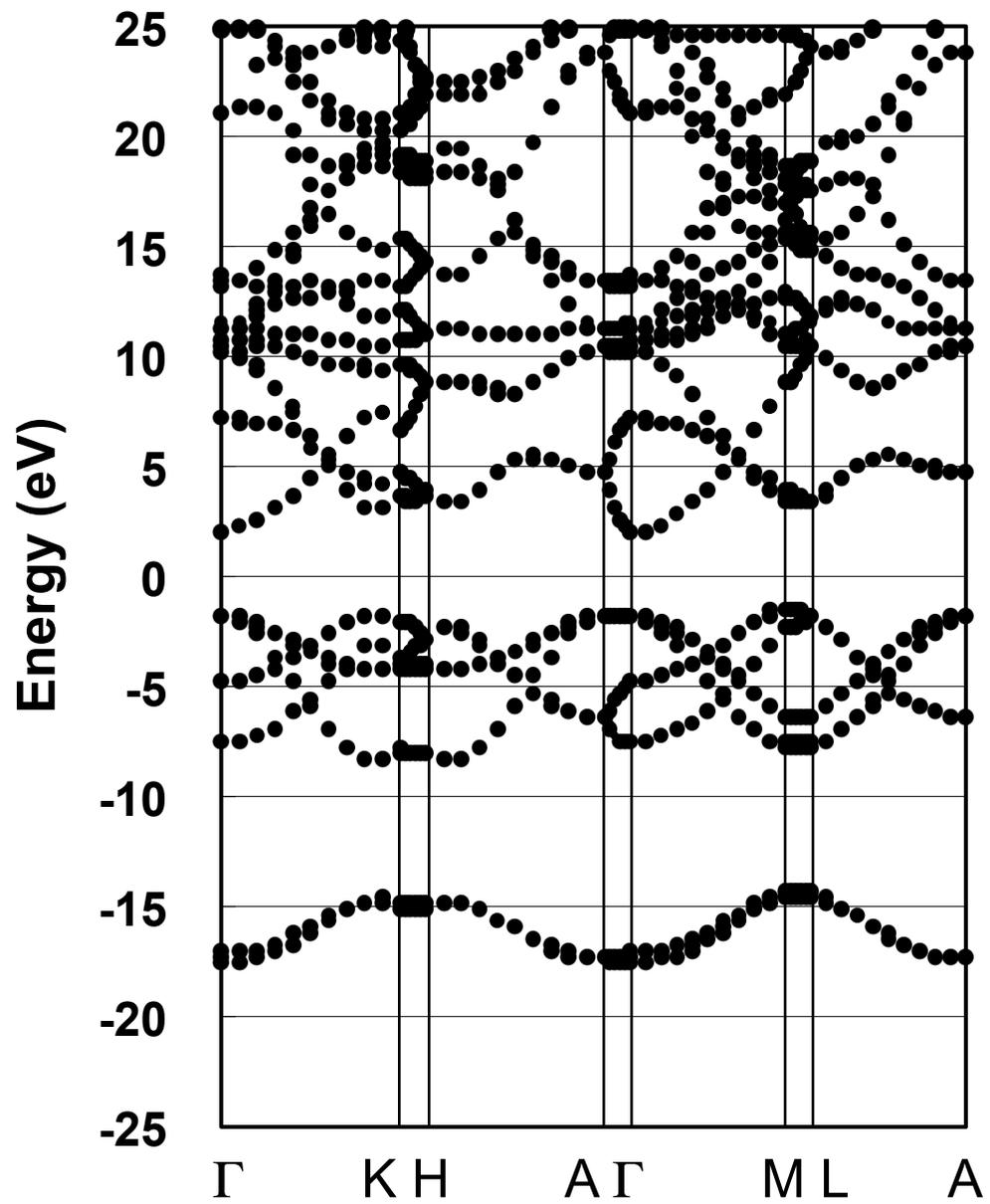

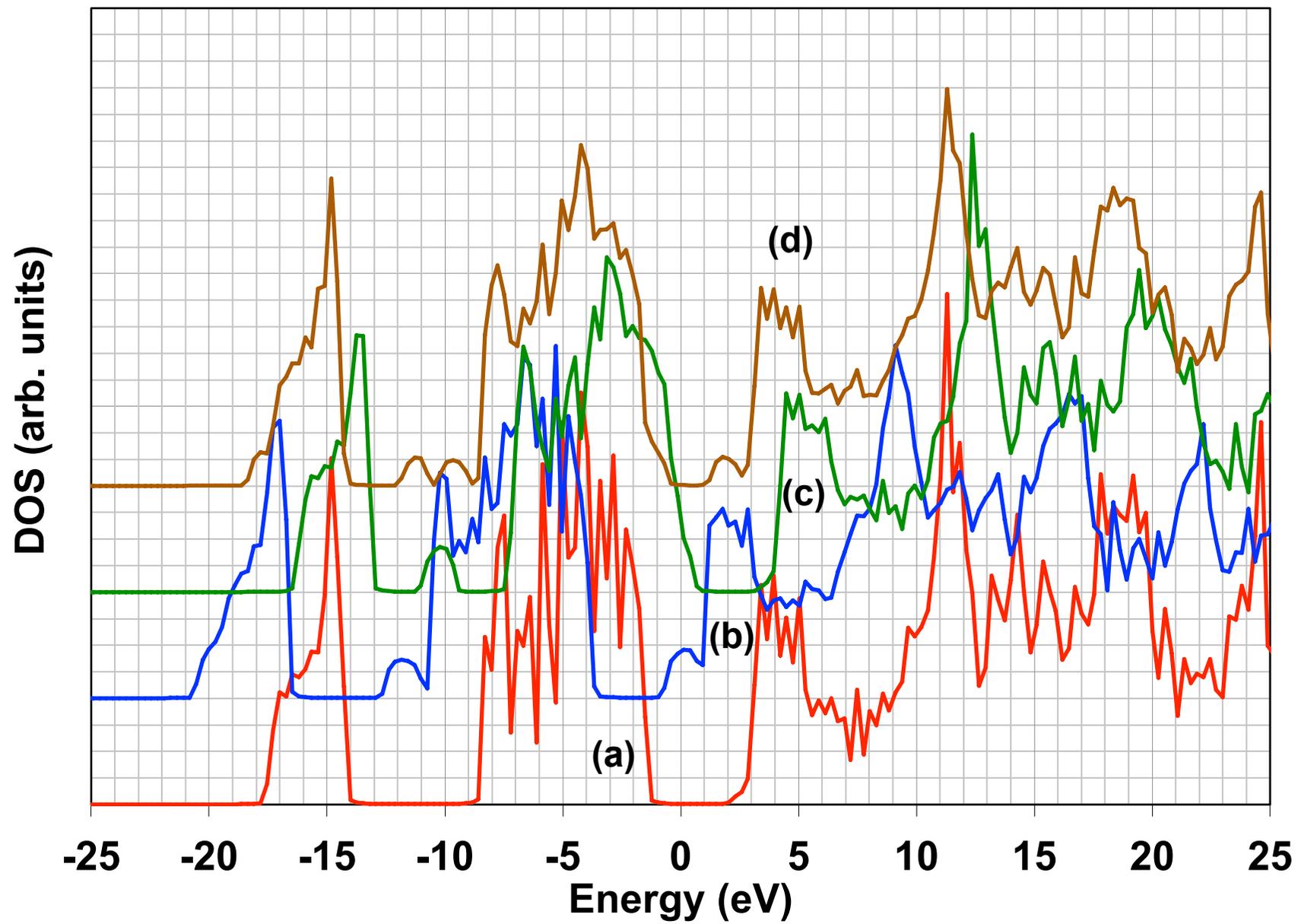

Fig. 15